\newcommand{\be}{\begin{equation}}
\newcommand{\ee}{\end{equation}}
\newcommand{\bea}{\begin{eqnarray}}
\newcommand{\eea}{\end{eqnarray}}
\newcommand{\ben}{\begin{eqnarray*}}
\newcommand{\een}{\end{eqnarray*}}
\documentclass{appolx}
\usepackage{epsfig}
\usepackage{amssymb}

\begin{document}
\title{Frustration-driven QPT in the 1D extended anisotropic Heisenberg model
\thanks{Presented at CSMAG'07 Conference, Ko\v{s}ice, 9-12 July 2007}}
\author{Evgeny Plekhanov, Adolfo Avella, and Ferdinando Mancini\\
\address{Dipartimento di Fisica "E. R. Caianiello" - Unit\`{a} CNISM di Salerno,\\
Universit\`{a} degli Studi di Salerno I-84081 Baronissi (SA), Italy }}
\maketitle
\begin{abstract}
By using Density Matrix Renormalization Group (DMRG) technique we study
the 1D extended anisotropic Heisenberg model. We find that starting from
the ferromagnetic phase, the system undergoes two quantum phase
transitions (QPTs) induced by frustration. By increasing the
next-nearest-neighbor (NNN) interaction, the ground state of the system
changes smoothly from a completely polarized state to a NNN correlated
one. On the contrary, letting the in-plane interaction to be greater
than the out-of-plane one, the ground state changes abruptly.
\end{abstract}
\section{Introduction}
QPTs, contrarily to ordinary ones (at $T\ne0$), are characterized by a
change of ground state on continuous varying the external
conditions~\cite{sondhi}. Such phase transitions are often caused by
frustration. One-dimensional quantum spin systems offer a wide variety
of examples of QPT. Among them, the Heisenberg-like models are the most
important. While 1D Heisenberg model with only nearest-neighbor
interaction possesses a well studied analytic solution by
Bethe anzatz~\cite{bethe}, the addition of NNN term suffices
to break the integrability. In such extended system a whole set of new
phenomena occurs. The physics of this model has been already largely
studied by means of many field-theoretical as well as numerical
methods~\cite{sorchub}, primarily in the proximity of the
antiferromagnetic phase.

However, the ferromagnet is also one of the ground states of the model
and it is primarily promoted by the diagonal part of the Hamiltonian.
If other terms in the Hamiltonian induce frustration, then eventually
the ferromagnet will "melt", and the system will undergo a QPT towards
some new phase.  In the present work we investigate such melting driven
by two distinct frustrating factors: i) the anisotropy of the nearest
neighbors interaction term and ii) the antiferromagnetic NNN
interaction.
\section{Model and Method}
We study the one-dimensional anisotropic
extended Heisenberg model with NNN interaction:
\be
   H = -J_z \sum_i S^z_i S^z_{i+1} + J_{\bot} 
   \sum_i ( S^x_i S^x_{i+1} + S^y_i S^y_{i+1})
     +  J^{\prime}\sum_i \mathbf{S}_i \mathbf{S}_{i+2}.
   \label{ham}
\ee
We find the ground state of the Hamiltonian (\ref{ham}) numerically by
means of the DMRG~\cite{dmrg} technique on a chain with 100 sites,
subject to open boundary conditions. We use $J_z>0$, which corresponds
to ferromagnetic coupling. In the present work we calculate the Fourier
transform of the correlation functions, both in-plane ($\langle
S^{x}(k)S^{x}(k+q)\rangle$) and out-of-plane ($\langle
S^{z}(k)S^{z}(k+q)\rangle$). Due to the large system size, the above
correlation functions do not depend on $k$, but only on the transferred
momentum $q$. The starting point for our analysis is the doubly
degenerate completely polarized state with all spins either ``up'' or
``down'', located on the phase diagram of the system within the limits
$0<J_{\perp}<J_z$ and $J^{\prime}\lesssim 0.31 J_z$, as found in
Ref.~\cite{us}.
\section{Frustration due to $J^{\prime}$}
Up to the values of $J^{\prime}\approx 0.31 J_z$ the system remains
totally polarized. In the momentum space this corresponds to a
$\delta-$function peak at $q=0$ in $z-$channel and a constant value of
$1/4$ in the perpendicular one. On the other hand, in the limit
$J^{\prime}\to\infty$ the system behaves as a couple of non-interacting
antiferromagnetic Heisenberg chains with nearest-neighbor interaction.
It is clear that in the case of $J^{\prime}\to\infty$ the system will
exhibit short-range AF correlations with exponential decay in real
space, which corresponds, in momentum space to a couple of Lorentzian
peaks centered at $q=\pm\pi/2$.

\begin{figure}[htp]
     \begin{center}
       \includegraphics[ width = 0.45\textwidth]{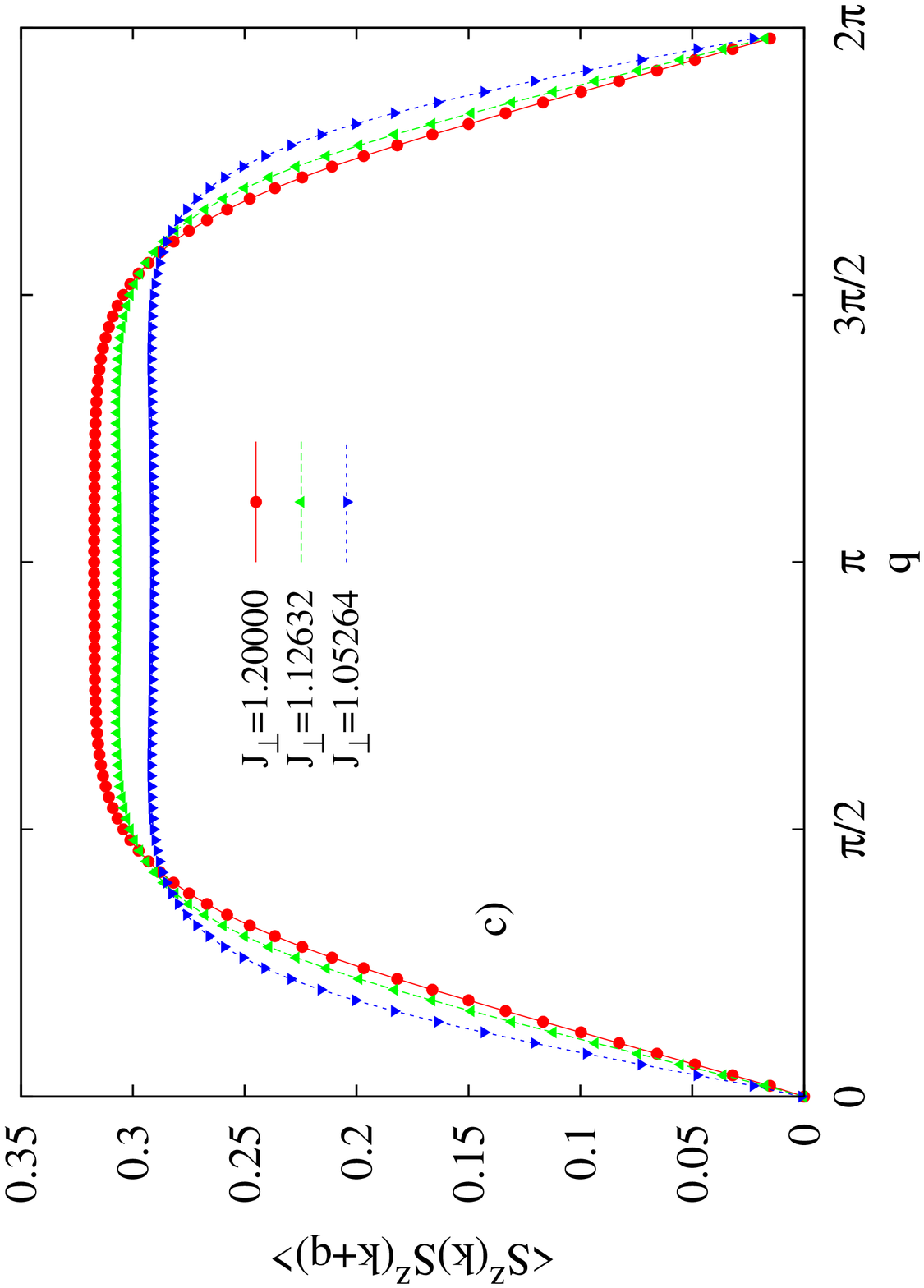}
       \includegraphics[ width = 0.45\textwidth]{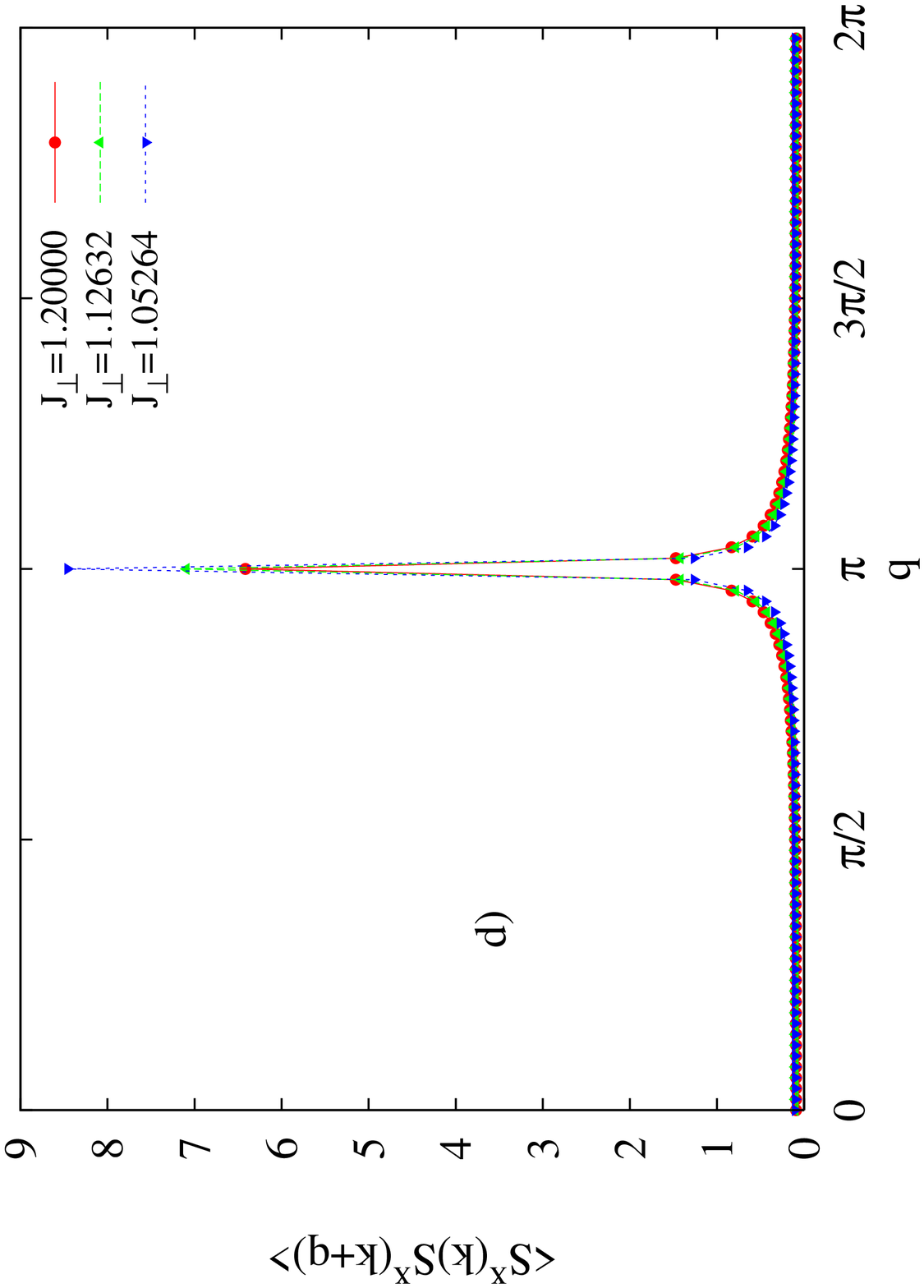}
       \includegraphics[ width = 0.45\textwidth]{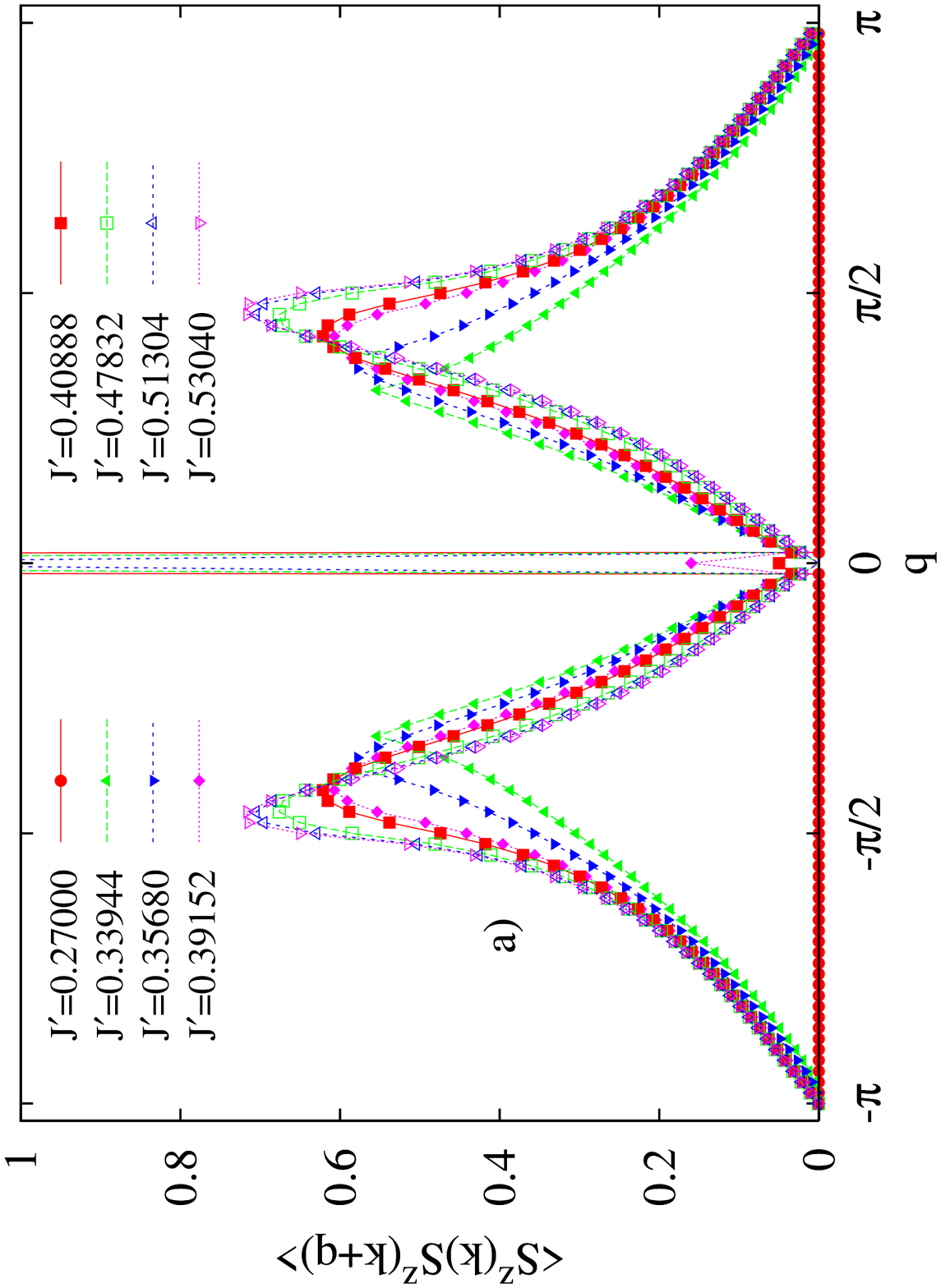}
       \includegraphics[ width = 0.45\textwidth]{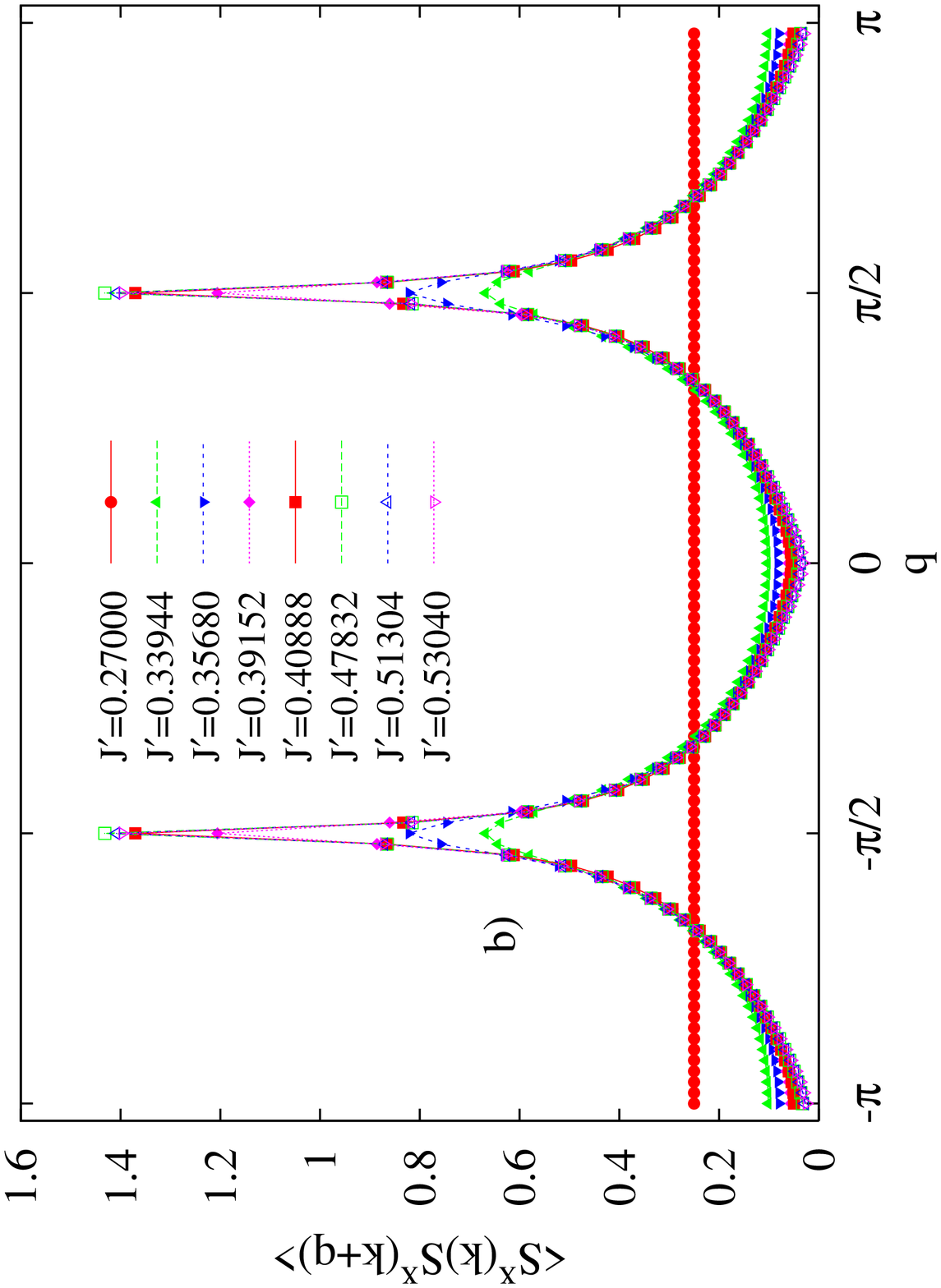}
       \caption{\small 
	   Momentum space spin-spin correlation functions for two
	   representative lines in $J^{\prime}-J_{\perp}$ plane both along
	   $z$ direction and in-plane: panel a) and b)
	   $J_{\perp}=0.2J_z$ and $J^{\prime}\in[0.27,0.53]$; panel c) and
	   d) $J^{\prime}=0.2J_z$ and $J_{\perp}\in[1.,1.2]$.
	   }
	   \label{fig}
     \end{center}
\end{figure}

Smooth change of $J^{\prime}$ causes a smooth rearrangement of the
ground state. Namely, for $J^{\prime} \gtrsim 0.31 J_z$ the system
abandons the totally polarized state, although conserving much of the
magnetization along the $z-$axis (see the peaks at $q=0$ on
Fig.~\ref{fig}a) for $J^{\prime} > 0.31 J_z$). At the same time a
short-ranged incommensurate phase appears: two strong Lorentzian-type
peaks develop in the vicinity of $q=\pm\pi/2$, the larger is
$J^{\prime}$ the closer are the peaks to $\pm\pi/2$.

On the contrary, in the perpendicular channel we have perfect
commensurability (at least for the values of $J_{\perp}\lesssim 0.5
J_z$) as shown on Fig.~\ref{fig}b).  As in the case of the $z$-axis, the
in-plane correlations appear to decay exponentially with the correlation
length depending on both $J_{\perp}$ and $J^{\prime}$.
\section{Frustration due to $J_{\perp}$}
Contrarily to the smooth change of the properties in the previous case,
frustration caused by $J_{\perp}$ changes the phase abruptly.  The
spins, totally polarized along $z$ direction for $J_{\perp}<J_z$, become
antiferromagnetically ordered in the $x-y$ plane (see Fig.~\ref{fig}c)). The
remaining $z-$component of the spins remain uncorrelated already at few
lattice spacings. In the limit $J_{\perp}\to\infty$ we expect
the system to become the $XX$ model, with an in-plane AF ground state.
As a matter of fact, at $J_{\perp}=1.2 J_z$ most of the correlations are
concentrated in the $x-y$ plane.
The scaling behaviour in the $x-y$ plane is once again exponential
(Lorentzian in $k$-space, see Fig.~\ref{fig}d)). The value of
$J^{\prime}$ does not qualitatively change this picture within the range
considered ($0<J^{\prime}<0.3 J_z$). Beyond this range the two
phases described above, should merge at some transition line, situated
somewhere in the region $J^{\prime}> 0.31 J_z$, $0.5 J_z<J_{\perp}
0.9 < J_z$.
\section{Summary}
In conclusion, by using a high-precision numeric tool (DMRG) we have
demonstrated how different types of frustration can lead to QPTs in the
1D extended anisotropic Heisenberg model. By considering a chain with
100 sites we have been able to observe the exponential decay of the
correlations and obtain a high resolution picture of the static spin-spin
correlations in momentum space. In contrast to the frustration induced
by $J_{\perp}$, the one due to $J^{\prime}$ leads to a
rather smooth rearrangement of the ground state. A more detailed study
of the two phases described in the present work and the transition
between them is in progress and will be reported elsewhere.

\end{document}